\documentclass[conference]{IEEEtran}
\IEEEoverridecommandlockouts
\usepackage{cite}
\usepackage{amsmath,amssymb,amsfonts}
\usepackage{algorithmic}
\usepackage{graphicx}
\usepackage{textcomp}
\usepackage{xcolor}
\usepackage{url}
\def\BibTeX{{\rm B\kern-.05em{\sc i\kern-.025em b}\kern-.08em
    T\kern-.1667em\lower.7ex\hbox{E}\kern-.125emX}}
\begin{document}

\title{/dev/SDB: Software Defined Boot – A novel standard for diskless booting anywhere and everywhere\\

\thanks{}
}

\author{
\IEEEauthorblockN{Aditya Mitra}
\IEEEauthorblockA{\textit{CyberMACS} \\
\textit{Kadir Has University}\\
Turkey \\
aditya.mitra@stu.khas.edu.tr \\
ORCiD: 0000-0002-9612-0810}
\and
\IEEEauthorblockN{Hamza Haroon}
\IEEEauthorblockA{\textit{CyberMACS} \\
\textit{Kadir Has University}\\
Turkey \\
hamzaharoon@stu.khas.edu.tr}
\and
\IEEEauthorblockN{Amaan Rais Shah}
\IEEEauthorblockA{\textit{CyberMACS} \\
\textit{Kadir Has University}\\
Turkey \\
amaanshah@stu.khas.edu.tr \\
ORCiD: 0009-0006-4453-6298}
\and
\IEEEauthorblockN{Mohammad Elham Rasooli}
\IEEEauthorblockA{\textit{CyberMACS} \\
\textit{Kadir Has University}\\
Turkey \\
m.rasooli@stu.khas.edu.tr}
\and
\IEEEauthorblockN{Bogdan Itsam Dorantes Nikolaev}
\IEEEauthorblockA{\textit{CyberMACS} \\
\textit{Kadir Has University}\\
Turkey \\
bnikolaev@stu.khas.edu.tr \\
ORCiD: 0009-0008-5119-1683}
\and
\IEEEauthorblockN{Tuğçe Ballı}
\IEEEauthorblockA{\textit{Department of Management Information Systems} \\
\textit{Kadir Has University}\\
Turkey \\
tugce.balli@khas.edu.tr}

}

\maketitle

\begin{abstract}
A computer is nothing but a device that processes the instructions supplied to it. However, as computers evolved, the instructions or codes started to be more complicated. As computers started to be used by non-technical people, it became imperative that the users be able to use the machine without having underlying knowledge of the code or the hardware. An operating system became the backbone for translating the inputs from the user to actual operation on the hardware. With increasing complexity and choices of operating system, it became clear that different groups of people, especially in an enterprise scenario, required different operating systems. Installing them all on a single machine, for shared computers became a difficult task, giving rise to network-based booting. But network-based booting was confined to only wired connectivity, keeping it restricted to very small geographical areas. The proposed system, /dev/SDB, is aimed at creating a standard where any user, can access the operating system authorized to them without having to be on the corporate network. It aims to offer the same over Wi-Fi as well as cellular connectivity, ensuring employees can truly work from anywhere, while following the policies for operating systems and without redundant hardware.
\end{abstract}


\section{Introduction}
Booting is usually one of the first steps after turning on the computer that loads an operating system. The operating system is, in turn, responsible for handling all activity of the computer. But it has been seen that there are various operating systems with different purposes and not everyone needs the same, especially in enterprise scenarios. Some components of the computer, like the storage, are often redundant and is only used to store the operating system. To boot a computer with persistent storage media, a standard for network boot is used. This usually makes use of Preboot Execution Environment (PXE). However, previous studies about PXE based deployment and network-booting in enterprise scenarios have been limited to wired networks, especially datacenter and enterprise networks. 

However, with the changing ways of working in modern society, a work-from-anywhere model is now adopted by various organizations. The employees mostly use remote systems, connecting from anywhere over the internet. Thus, the previous consideration of wired connectivity is becoming less relevant in the modern workforce. While thin clients, and VDI (Virtual Desktop Infrastructure) are being implemented to overcome security concerns and to keep all operations running from a centralized system, it often is not enough from both a security perspective as well as from a financial perspective. Users often use general purpose operating systems booting from a persistent storage media which threaten risks involving malware, viruses, etc. Further a general-purpose operating system might be redundant for all employees since not everyone would use every feature of the operating system in their daily workflow. The persistent storage media are usually redundant and only used to hold the operating system and open an attack surface. Users with general-purpose operating systems often tend to install external applications on the computer system which threatens the confidentiality of enterprise resources. 

In modern office-like environments, even if the computer systems have wired connectivity, any employee from any department is usually free to choose any desktop and work at any time. This is usually not limited to a single site; the employees may often travel to different locations or offices of the organization and get to work. Different employees often use different operating systems, for example a developer may use a Linux distro like Ubuntu, a pentester may use a distro like Kali, Parrot or Black Arch, a person in sales could use Zorin, tailored for meetings, while one in accounting may use a custom operating system only tailored to spreadsheets, and other required softwares. This demands the systems be dynamically provisioned tailored to the requirements of each employee based on their role right from the operating system level. 

The proposed standard is aimed at developing a system where a computer system, including wireless and portable computers, could be booted into any one of multiple operating systems, diagnostic systems or any portable operating system-based environment (like WinPE) depending on user permissions and user requirements. It deals with all the problems and offers a system to authenticate the user even before the system is booted. The system then retrieves the operating system assigned to the user and boots up using the target image. It uses in-memory booting, often loading the entire OS image from a Network File Share (NFS) over the internet. The system offers connectivity to the boot server over wired, wireless and cellular network, preferably even satellite network in future, to facilitate secure work-from-anywhere paradigm for the users while ensuring the user always gets the operating system tailored to his requirements.

The paper is divided into the following sections: Section 1 deals with the introduction, section 2 deals with the literature review, section 3 deals with the methodology of the study, section 4 represents the experimental setup used in the study, and section 5 is the threat model.

\section{Literature Review}

The proposed system encompasses various concepts related to how an operating system is loaded. This includes bootloaders, Preboot execution environment (PXE), a few open-source extensions of PXE like iPXE and gPXE, followed by existing studies on diskless cloud based booting systems.

Bootloaders form the foundation of the operating system startup process, initializing hardware and loading the kernel after system firmware execution. Traditional multi-stage designs, as described by Sebastian and Sankar \cite{b1}, relied on small first-stage loaders in the Master Boot Record (MBR) to locate secondary loaders like GRUB or Syslinux, which provide configuration menus and multi-boot capabilities. The shift from BIOS to the Unified Extensible Firmware Interface (UEFI) introduced secure boot features, enforcing digital-signature validation before kernel execution to prevent unauthorized firmware \cite{b2}. 

Modern bootloaders, such as GRUB 2 and Syslinux/PXELinux, incorporate chainloading and modular configuration to support diverse boot sources, including disks and networks \cite{b3}, \cite{b4}. However, research shows that memory-safety flaws during the pre-kernel stage can undermine system trust, emphasizing the need for verified, secure boot code \cite{b5}. Recent developments extend bootloaders beyond local environments: Park et al. proposed FLEX-IoT \cite{b6}, which uses hash-based authentication and adaptive transfer scheduling for secure network booting in IoT devices, while Mitra et al. introduced Colaboot \cite{b7}, a diskless cloud-booting model leveraging network protocols like Dynamic Host Configuration Protocol (DHCP) and Trivial File Transfer Protocol (TFTP) to virtualize the OS loading process. 

The idea that network booting has traditionally simplified OS deployment, centralized management, and removed reliance on local storage is backed by several studies researched over time. The research highlights \cite{b8} that a PXE (Pre-boot Execution Environment) boot system is a network protocol enabling a computer to boot and install an operating system over a network. It serves as an effective process that streamlines the installation process and diminishes the likelihood of errors and inconsistencies, especially in large organizations where many computers require the same OS and software setup. The role of PXE standardization is detailed in Johnston and Venaas' 2006 RFC \cite{b9}. It specifies DHCP options used by PXE and EFI clients to uniquely identify booting machines and their pre-OS environments, enabling the DHCP and PXE servers to deliver the appropriate OS bootstrap images.

The reliance on wired connections is implied throughout the papers but is most evident in research on embedded systems. For example, U-Boot on ZUS+ devices enables network booting via the Preboot Execution Environment (PXE). The initial boot steps in the standard ZUS+ process include application-specific details that reflect physical infrastructure needs. Fuchs et al.'s Split Boot \cite{b10} research proposes methods to shift all application-specific settings to a network storage device, which is then automatically accessed during boot, although wired connectivity remains necessary.

The Trivial File Transfer Protocol (TFTP), commonly used for network booting, has notable security flaws and performance issues. During network boot, the server faces high demand due to numerous IoT devices requesting system images based on their different roles, resulting in a substantial network load \cite{b6}. When multiple boot requests co-occur, the server's resource usage increases exponentially.

The DHCP coupling limitation is apparent in the core architecture outlined in various papers. The Network Boot System (NBS) relies on three main protocols: PXE, DHCP, and TFTP, showing a close interconnection. During the process, the PXE-enabled client must request an IP address via PXE and DHCP, establishing reliance on the DHCP system. \cite{b11}

Performance studies document the scalability challenges faced by large-scale fleets. Research indicates that transmitting numerous system images to IoT devices within an IoT platform leads to high network traffic.\cite{b6} As the demand for transmitting system images grows, it will further increase network traffic consumption. 

The iPXE boot firmware is recognized as an extension of traditional PXE, offering capabilities beyond legacy PXE/TFTP. Stites et al. note that iPXE enables booting via HTTP and iSCSI (in addition to TFTP), with flexible scripting to orchestrate complex boot processes \cite{b12}. 

Chainloading is a feature wherein a minimal PXE ROM hands off control to iPXE, allowing existing PXE infrastructure to leverage iPXE’s enhancements without changing NIC firmware. Unlike standard PXE, iPXE can embed CHAP credentials in boot scripts, allowing diskless clients to securely boot from iSCSI targets without storing secrets on the NIC. Recent research has integrated iPXE into secure and scalable architectures; Oliveros et al. extended iPXE with a BitTorrent-like peer-to-peer protocol to distribute boot images and eliminate the single-server bottleneck \cite{b13}. 

Moser demonstrates a stateless iPXE-based boot environment with UEFI Secure Boot and script signing, ensuring end-to-end boot integrity \cite{b14}. However, iPXE’s built-in TLS support tops out at TLS 1.2 with deprecated ciphers, requiring careful configuration to maintain security. Moser also experimented with embedding iPXE in NIC firmware to replace vendor PXE code; while this gave a seamless boot path, it proved impractical at scale due to device-specific firmware tooling. In cloud data centers, iPXE serves as the network boot layer for bare-metal provisioning, prompting researchers to address trust implications. Schear et al. present a cloud bootstrapping architecture leveraging TPM/vTPM to securely provision physical and virtual machines \cite{b15}. For instance, M2 (Malleable Metal as a Service) uses iPXE to orchestrate diskless provisioning of multi-tenant physical servers: M2 chainloads iPXE on each boot to bypass NIC limitations and load OS images via HTTP or iSCSI, treating bare-metal nodes as cloud instances \cite{b16}.

Cloud-based booting is the process of booting systems from the cloud, which does not require local storage for scalability and security \cite{b17}. Colaboot proposes machines to remotely download only a minimal bootloader and stream the OS from the cloud, executes the OS within the volatile memory to ensure consistency and resilience \cite{b7}. Diskless Remote Booting systems like ClassCloud boot off a central server using protocols like PXE and TFTP, addressing largescale enterprise needs but face challenges in heterogeneous hardware environments \cite{b18}. Another academic approach proposed a multi-step bootstrapping system that integrates cloud infrastructures to manage the workflow flexibly \cite{b19}. 

Proprietary solutions include Dell's BIOSConnect, which allows for cloud-based recovery by downloading an OS image into RAMDisk when the local recovery partition is not present to ensure system recovery \cite{b20}. Microsoft's Windows 365 Boot enables users to directly boot into their cloud- hosted "Cloud PC", a solution for shared-device environments \cite{b21}. While this approach finds its prime application in optimized power, scalability, and management, it may pose network latency challenges in diskless environments.

The extensive literature review highlights the research gaps that it is the first study to use software to define operating systems to be used per user in an enterprise scenario, while at the same time allowing for remote booting over any form of connection, wired, wireless, cellular, etc. The system aims to integrate a hardware module to the computer system to be able to achieve this.

\section{Methodology}

The proposed standard offers a system to boot multiple systems, regardless of the location and connectivity type to the operating system provisioned by the administrator to the user.

The system is built in two different parts: 
\begin{itemize}
    \item	A hardware module to be able to handle network connectivity before booting
    \item	A cloud module to handle user authentication and OS images.
\end{itemize}

\subsection{Hardware Module}
The hardware module to handle network connectivity is based on a lightweight Single Board Computer (SBC) running a lightweight distro like Alpine Linux. The module has two network interfaces, one of which is to be connected to the host computer internally via PCIe or via other physical media like a short UTP connector. The second interface is connected to upstream connectivity like a Wi-Fi adaptor, a cellular modem or a wired connector.

The module bridges both interfaces. If a DHCP server is discovered on the upstream network, it enables a Proxy DHCP server on the bridge interface. If a DHCP server is not discovered, the module sets up a static IP address on the bridge interface and starts DHCP server. Further, the module hosts a TFTP server serving a custom iPXE based bootloader image for the target computer which attempts to chain load from a given web server addressed by domain name.

Both the proxy DHCP server and the standard DHCP server, enabled based on the scenarios, points to the iPXE based bootloader for the boot-file and its own IP for the next-server options for DHCP. The standard DHCP server offers the module’s own IP for DNS server, while the proxy DHCP server does not change the DNS configuration offered by the upstream DHCP server. The iPXE based bootloader chain loads a boot file from the cloud module, pointed by the domain name.

The hardware module hosts a Flask based webserver module that is used to configure the Wi-Fi and Cellular connectivity. The webserver module returns responses in the form of iPXE forms and menus which could be rendered by the target PC even without an operating system.

The hardware module runs a DNS server that resolves all names to its own address without any recursive resolution from upstream. When the target computer is turned on, it enters PXE environment, sending an DHCP Broadcast. If the upstream network is not connected or it does not have a DHCP server, the hardware module returns a DHCP Offer with an IP in the same subnet as its static IP. It looks for the next-server field and finds the TFTP server and attempts to load the iPXE based bootloader. Once the bootloader is loaded, it attempts to chain load from the domain name of the cloud.

However, since the DNS server supplied points to the hardware module itself, it resolves to the hardware module, and the flask-based application is triggered. It serves iPXE forms requesting the Wi-Fi SSID, Password, or other cellular network configuration information. Once it is connected to the internet, the same process repeats.

Now when it is connected to the internet on the upstream, the proxy-DHCP server allows the target machine to use IP address from the upstream network itself. It still uses its own IP for the TFTP server and serves the same bootloader for chain loading.

\begin{figure}[htbp]
    \centering
    \includegraphics[width=\linewidth]{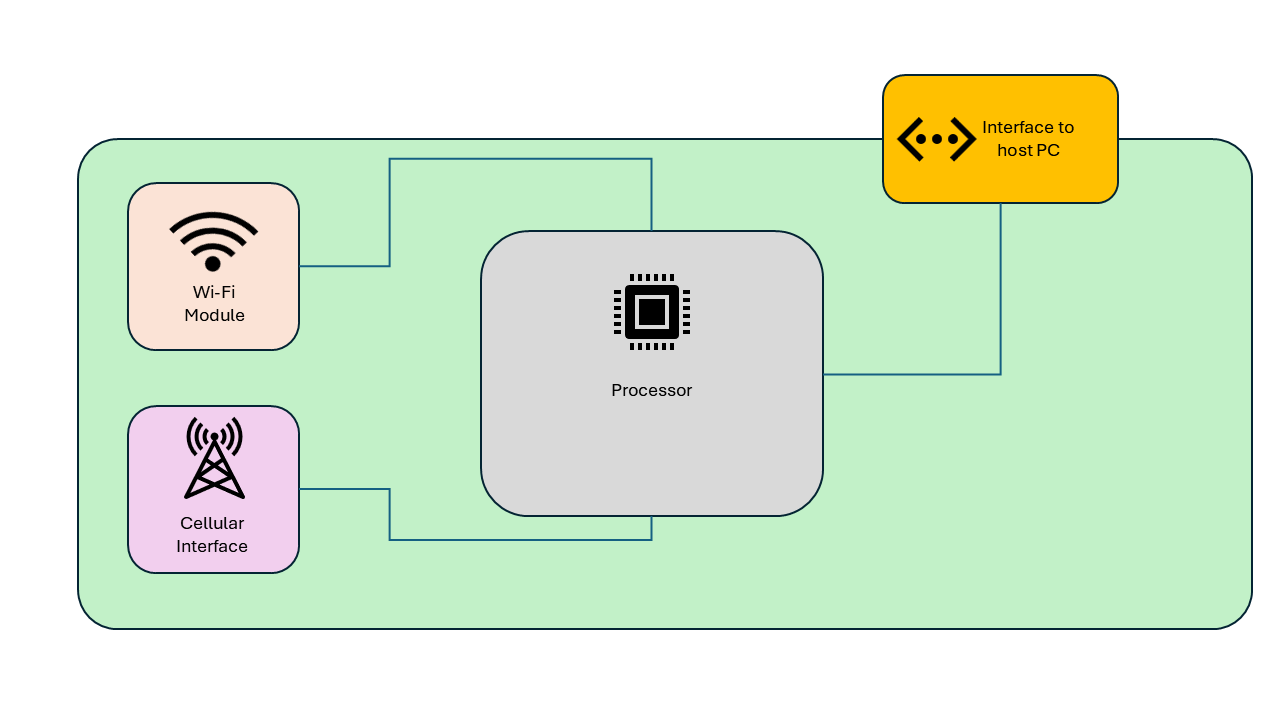}
    \caption{Schematic of Hardware Module}
    \label{hardwareschematic}
\end{figure}

Once the DNS server resolves the actual cloud server by the domain name, the boot-loader chain loads from the cloud, which first presents a form for Username and Password of the user. The cloud module authenticates the user based on the username and password, followed by serving the operating system assigned to the user. Figure \ref{hardwareschematic} shows the schematic diagram of the hardware module.

\subsection{Cloud Module}

The cloud module consists of a web server, a database, file storage and an optional Network File Share (NFS) server. It has two submodules:
\begin{itemize}
    \item Configuration module/ Admin module
    \item Usage module
\end{itemize}

\begin{figure}[htbp]
    \centering
    \includegraphics[width=\linewidth]{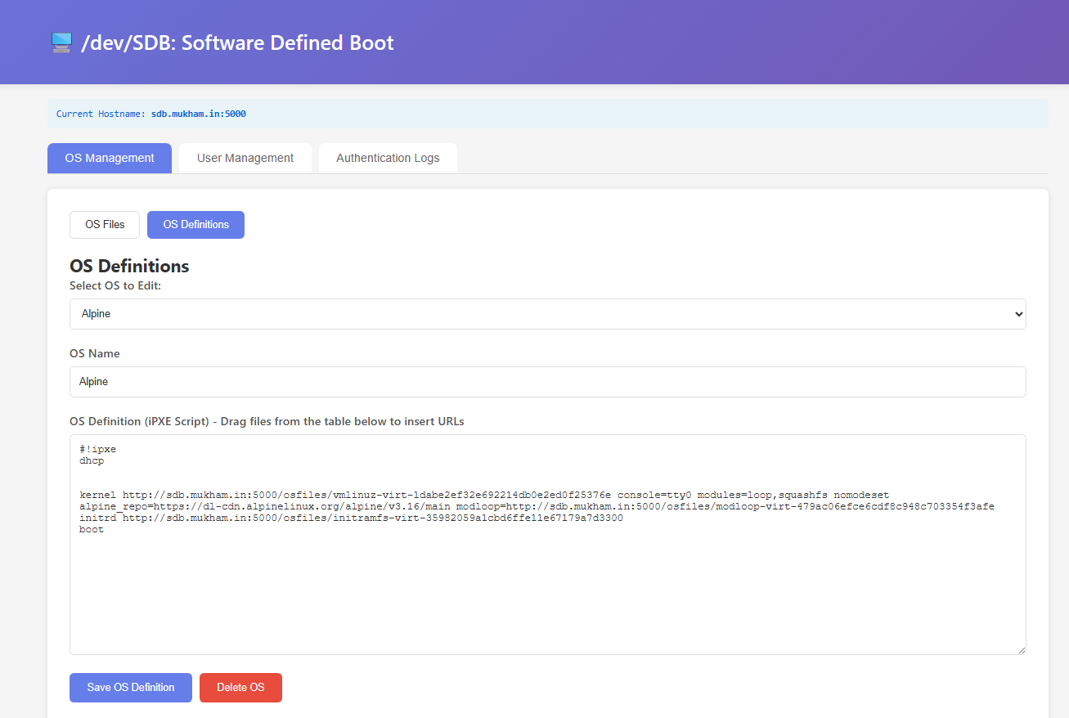}
    \caption{OS Definitions}
    \label{osdefinition}
\end{figure}

The configuration module behaves as a web application and allows an administrator to define operating systems and assign them roles. An operating system is defined by uploading the required files, (for example, the kernel and initial ram disk) and defining the kernel parameters in the format of a script. The configuration module also allows user management, including creating users, deleting users and assigning them operating systems. The OS Metadata, and user details are stored in an SQLite Database while the OS files are stored as files. The cloud module further logs all successful and unsuccessful authentication attempts, against the MAC addresses of the computer used and displays them in the web application. Figure \ref{osdefinition} shows how the Operating System is defined in the cloud module.

\begin{figure}[htbp]
    \centering
    \includegraphics[width=\linewidth]{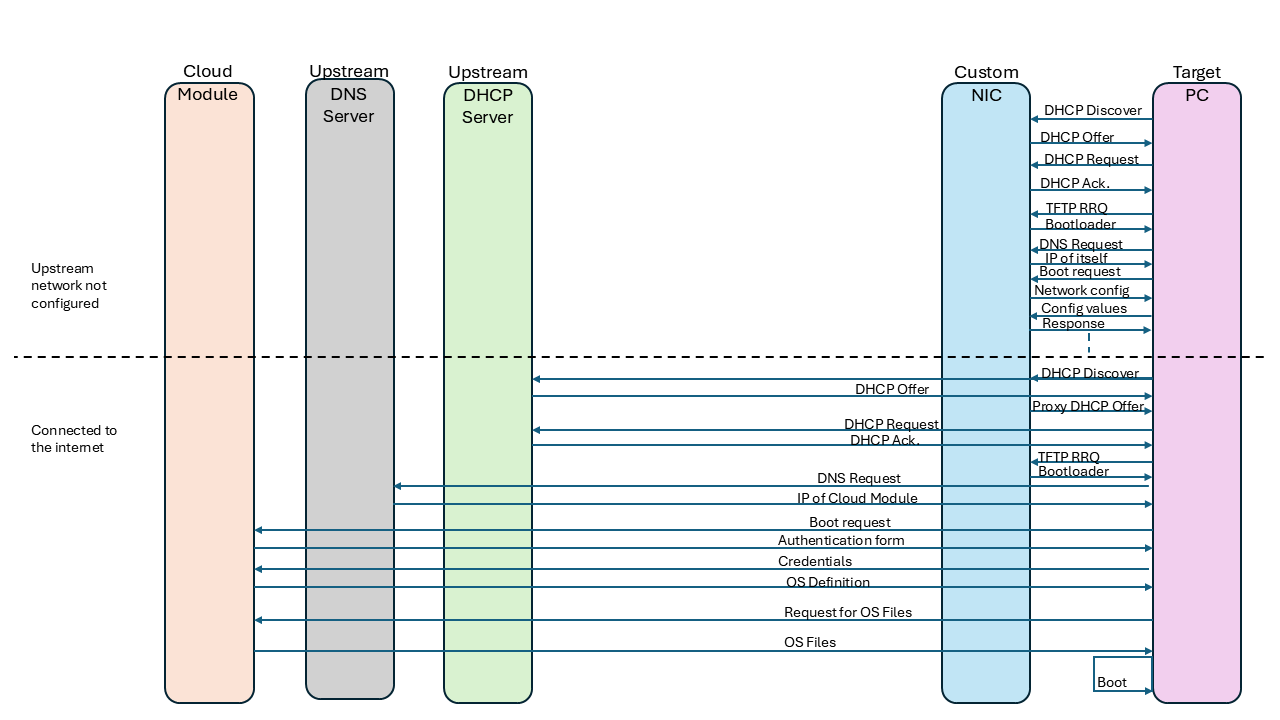}
    \caption{Workflow of the system}
    \label{workflow}
\end{figure}

Hence, the system allows having multiple users, and to assign operating systems to each user. Each user would be able to boot into an enterprise computer only after authentication. Further, it is easy to monitor which employee used which computer by the MAC Address and this gives further insights into the usage patterns of an enterprise. The system offers seamless onboarding and offboarding of users and offboarded users will be unable to boot into the enterprise system. Figure \ref{workflow} shows the workflow of the system.

\section{Experimental Setup}

The system was deployed in a simulation on GNS3 application. The Hardware Module NIC was developed as a template on top of an Alpine Linux image, with 128MB of RAM, ensuring it had as minimal footprint as possible. It was connected to the target PC, which was a standard QEMU appliance with no storage media and around 256MB of RAM. While in practice the target PCs will have a greater amount of memory, 256MB was ideal for the simulation. Three instances of target PCs were created to simulate different PCs in an enterprise environment. 

\begin{figure}[htbp]
    \centering
    \includegraphics[width=0.8\linewidth]{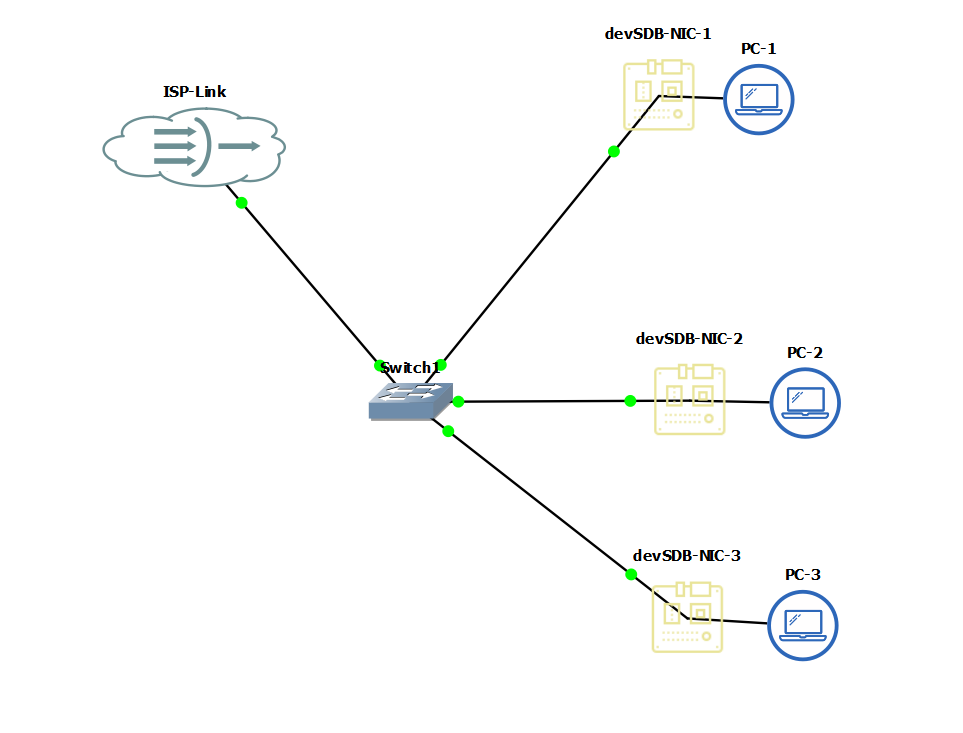}
    \caption{Experimental Setup}
    \label{expsetup}
\end{figure}

The cloud module was run on the host PC, outside the GNS3 simulation which automatically created a segmented network. This further ensured the target PC and the cloud module were not in the same broadcast domain and that the target PC was inside a GNS3 NAT, thus simulating real ISP CG-NAT environments. Figure \ref{expsetup} shows the experimental setup.

The cloud module was developed in Python with the Flask framework act as a web application and it uses SQLite to store user information and OS metadata. The uploaded files were stored in a directory by the flask application.

\begin{figure}[htbp]
    \centering
    \includegraphics[width=0.8\linewidth]{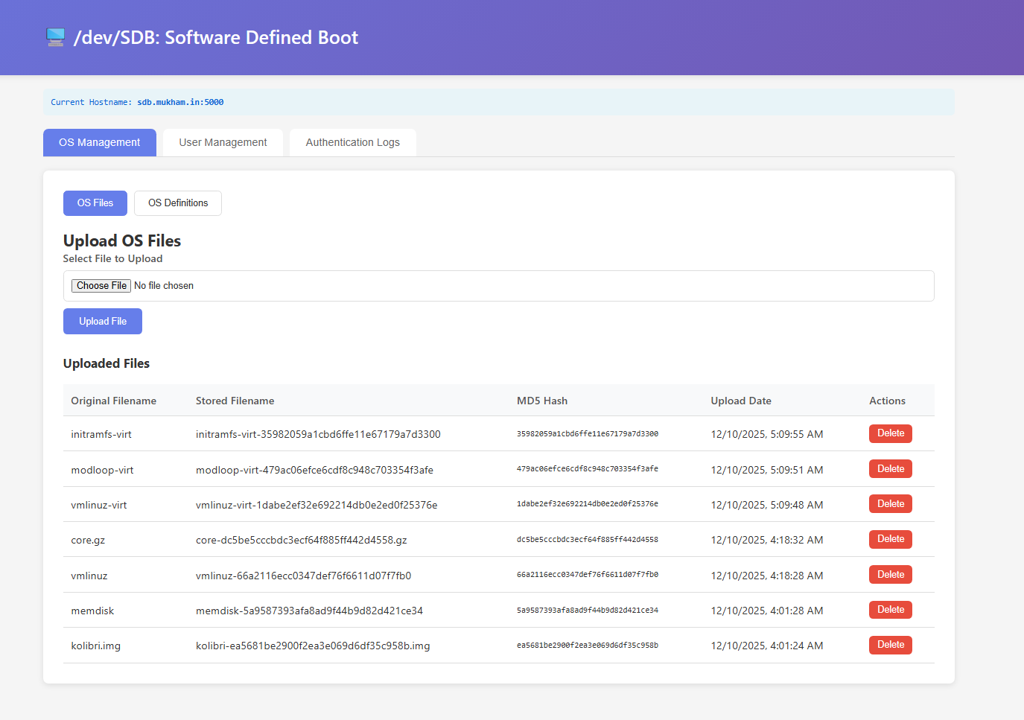}
    \caption{OS Files}
    \label{filemgmt}
\end{figure}

Three lightweight operating systems were defined from the admin module: Kolibri OS, Tiny Core Linux and Alpine Linux. The files of the operating system, including the kernel, initial ram disk image, and filesystem were uploaded to the cloud module. Figure \ref{filemgmt} shows the file management process of the cloud module.

\begin{figure}[htbp]
    \centering
    \includegraphics[width=0.8\linewidth]{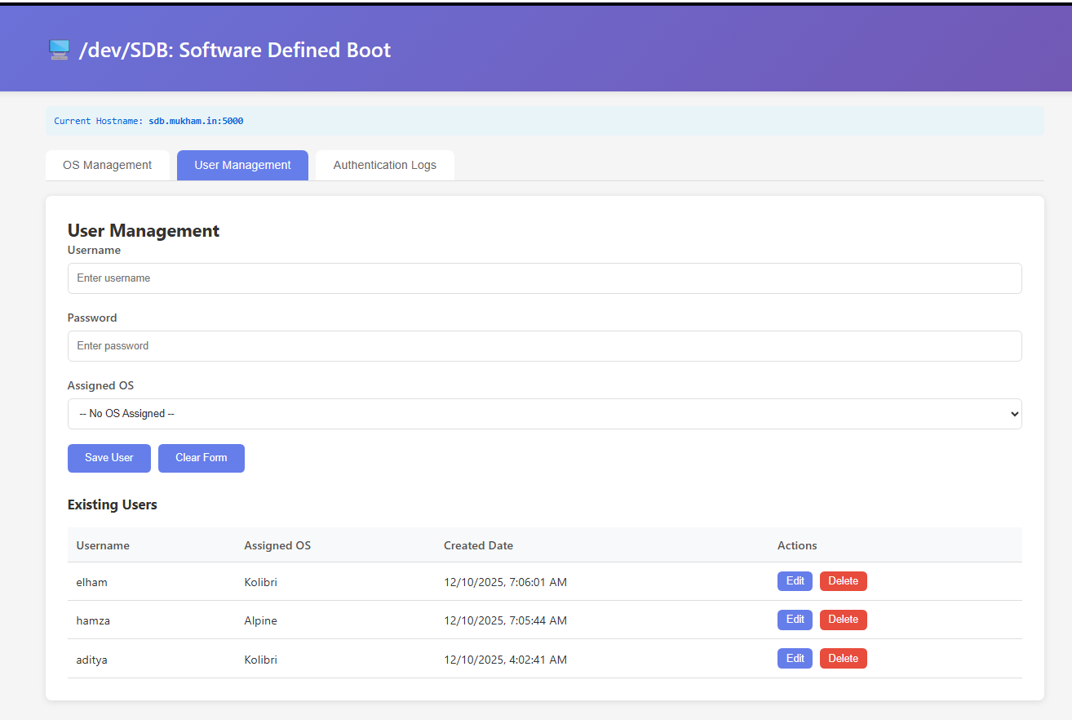}
    \caption{User Management}
    \label{user}
\end{figure}

Then, three users were created and assigned the operating systems to test. Figure \ref{user} shows the User management screen.

\begin{figure}[htbp]
    \centering
    \includegraphics[width=0.8\linewidth]{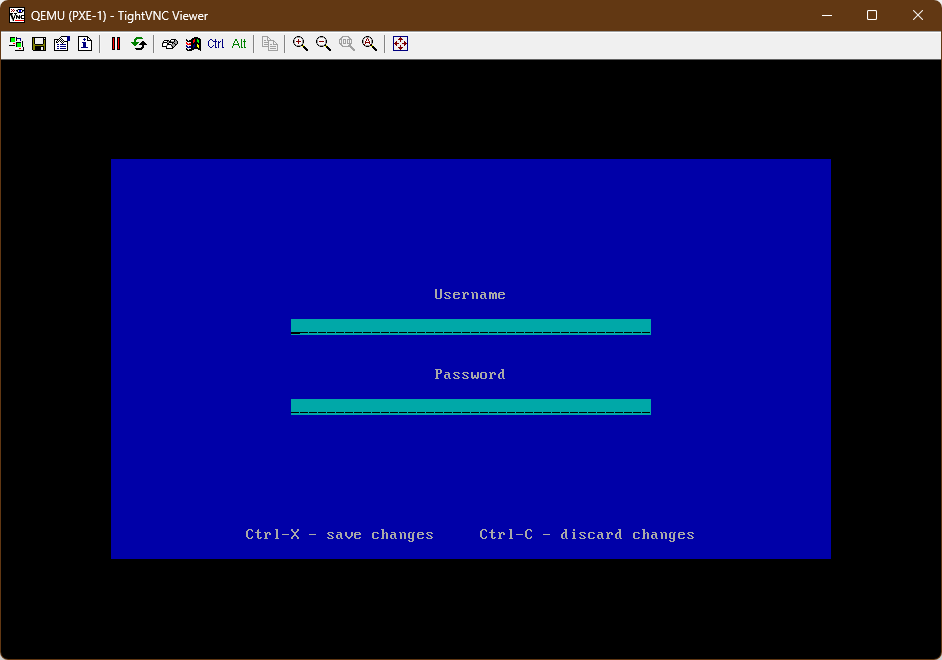}
    \caption{Login prompt}
    \label{login}
\end{figure}

The users were able to turn on the target PCs in the simulation and were presented with the login screen. The users entered their credentials and were booted to their assigned operating system automatically. Figure \ref{login} shows the login prompt on the target PCs.

\begin{figure}[htbp]
    \centering
    \includegraphics[width=0.8\linewidth]{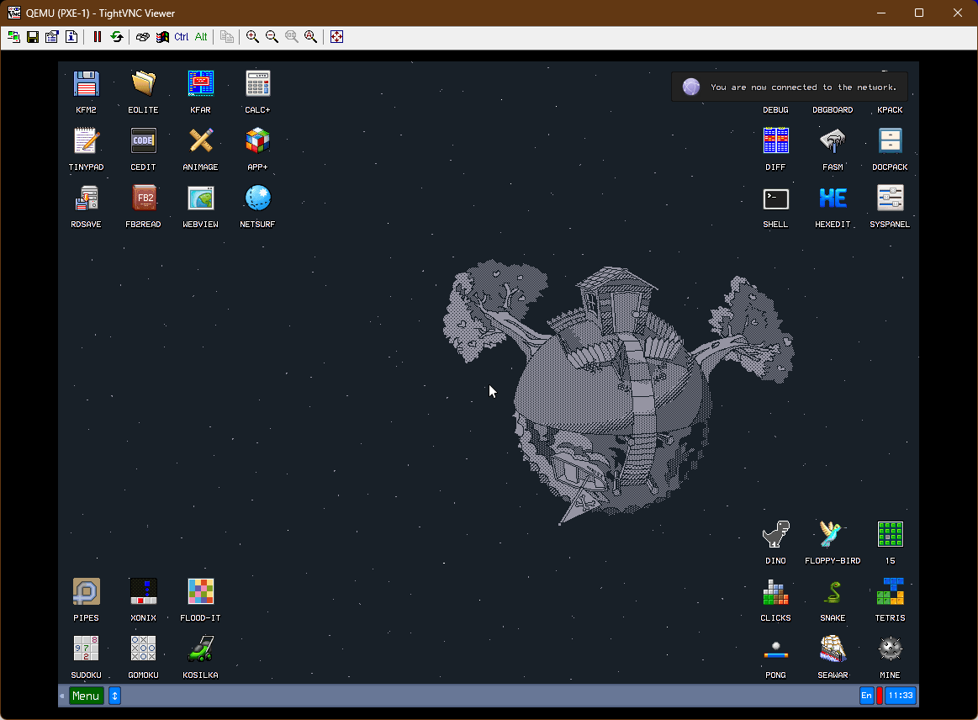}
    \caption{Kolibri OS Desktop}
    \label{kolibri}
\end{figure}

The target machine retrieved the OS files from the cloud module and booted, presenting the user the desktop or the CLI based on the operating system. Figure \ref{kolibri} shows the desktop of Kolibri OS after a successful boot.

\begin{figure}[htbp]
    \centering
    \includegraphics[width=0.8\linewidth]{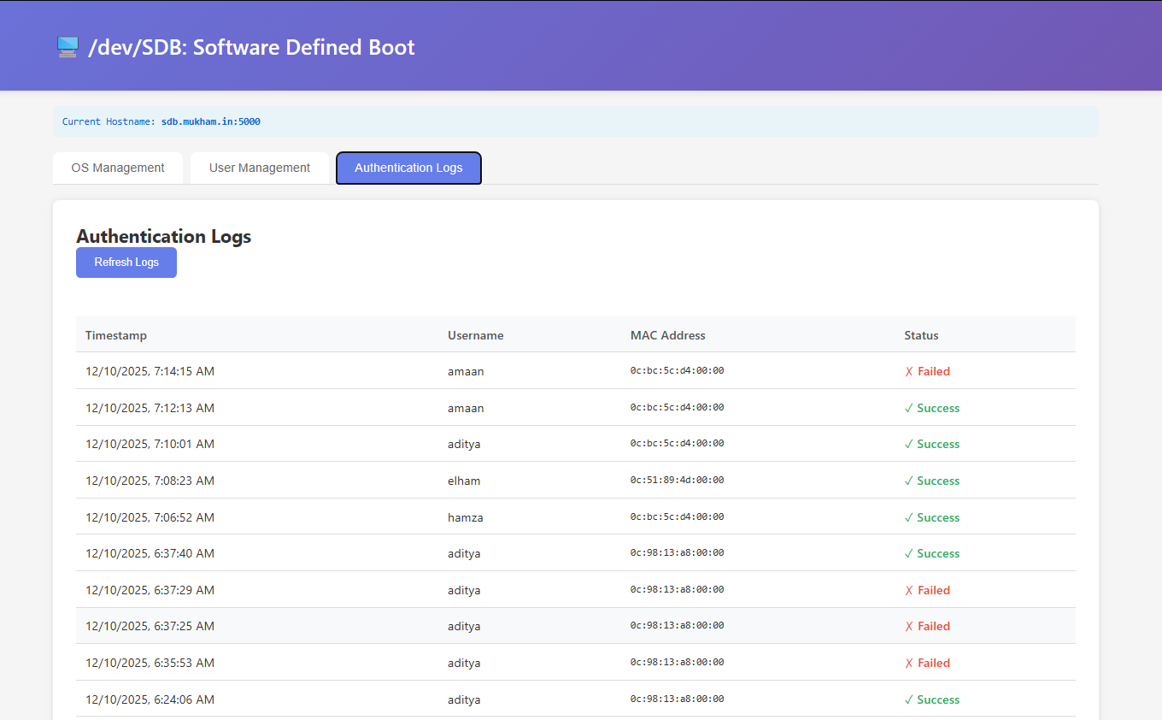}
    \caption{Authentication logs}
    \label{log}
\end{figure}

All user authentications, both successful and unsuccessful, were logged along with the MAC addresses of the target device. Figure \ref{log} shows the authentication logs.

Hence, the system was successfully deployed in a simulation and tested, giving consistent results. This study is not reporting the speed and performance of the setup deliberately because this is a simulation under ideal conditions and each boot took under 3 seconds. However, in real world conditions, the boot process may be affected by network, memory and other conditions.

\section{Threat Model}

The proposed standard aims to be more secure than existing PXE and other network booting standards. The threat model includes:
\begin{itemize}
    \item Proxy DHCP Poisoning: In standard PXE systems, a rogue machine in the broadcast domain may act that a rogue Proxy DHCP server, thus issuing malicious next-server (TFTP Server) IP addresses to the machines using PXE. In the proposed standard, the proxy DHCP is handled by the hardware module which is connected inside the target system. Hence, this is mitigated.
    \item Malware, including boot sector: Not having persistent storage media on the target PCs ensure that any malware affecting the system would be immediately flushed out from memory as soon as the computer is shut down.
    \item Access to unauthorized operating systems: Usually computers that have multiple operating systems installed often let a user use any. The proposed system, on the other hand, only serves the operating system the user is authorized to use, after authentication.
\end{itemize}

\section{Conclusion and Future Scope}

/dev/SDB promises to be a solution for embracing software defined operating system assignment to different users based on roles in an enterprise, remote environment. To establish /dev/SDB as a viable solution, extensive simulations have been performed on suitable software, emulating target PCs with minimal capabilities. All tests have resulted in successful operations.

The future scope of the study includes developing the proposed hardware module physically, testing it outside a simulation and stress testing the same in enterprise environments. Further the cloud module can be deployed in any public cloud or private cloud depending on the policies of the enterprise. 

The advancements proposed in /dev/SDB would be instrumental in reducing the use of redundant hardware in computers, while at the same time ensuring each user is able to be on an operating system tailored to them.

\section{Acknowledgement}

This research was conducted within the framework of the Erasmus Mundus Joint Master’s Programme in Applied Cybersecurity (CyberMACS) — Project No. 101082683, co-funded by the European Union under the Erasmus+ MUNDUS Programme.

\end{document}